\documentclass[10pt,conference]{IEEEtran}
\IEEEoverridecommandlockouts
\usepackage{cite}
\usepackage{amsmath,amssymb,amsfonts}
\usepackage{algorithmic}
\usepackage{graphicx}
\usepackage{textcomp}
\usepackage{xcolor}
\usepackage{tabularx}
\usepackage{comment}
\usepackage{multirow}
\usepackage{makecell}
\usepackage{array}
\usepackage{tikz} 
\usepackage{enumitem} %% Jianjun (20240330)
\usepackage{tabularray}
\usepackage{threeparttable}
\usepackage{makecell}
\usepackage{hyperref}
\usepackage{balance}
\usetikzlibrary{quantikz2}
\def\BibTeX{{\rm B\kern-.05em{\sc i\kern-.025em b}\kern-.08em
    T\kern-.1667em\lower.7ex\hbox{E}\kern-.125emX}}

\newcommand{\black}[1]{\textcolor{black}{#1}}

\newcommand{\tabincell}[2]{\begin{tabular}{@{}#1@{}}#2\end{tabular}}
\definecolor{ccr}{RGB}{10,110,150} 

\hypersetup{hypertex=true,
    colorlinks=true,
    linkcolor=ccr,
    anchorcolor=ccr,
    citecolor=ccr}

\begin{document}

\title{Quantum Circuit Ansatz: Patterns of Abstraction and Reuse of Quantum Algorithm Design
}
%\iffalse
\author{\IEEEauthorblockN{Xiaoyu Guo}
\textit{Kyushu University}\\
%Fukuoka, Japan \\
guo.xiaoyu.961@s.kyushu-u.ac.jp
\and
\IEEEauthorblockN{Takahiro Muta}
%\IEEEauthorblockA{%\textit{dept. name of organization (of Aff.)} \\
\textit{Kyushu University}\\
%Fukuoka, Japan \\
prunus.dr1293@gmail.com
%email address or ORCID
%}
\and
\IEEEauthorblockN{Jianjun Zhao\IEEEauthorrefmark{1}\thanks{\IEEEauthorrefmark{1} Corresponding Author. This work is supported in part by JSPS KAKENHI Grant No. JP23H03372, No. JP24K14908, and No. JP24K02920. The work is also supported in part by JST-Mirai Program Grant No. JPMJMI20B8.}}
\textit{Kyushu University}\\
%Fukuoka, Japan \\
zhao@ait.kyushu-u.ac.jp
}
%}
%\fi

\maketitle
\thispagestyle{plain} 

\begin{abstract}
Quantum computing holds the potential to revolutionize various fields by efficiently tackling complex problems. At its core are quantum circuits, sequences of quantum gates manipulating quantum states. The selection of the right quantum circuit ansatz, which defines initial circuit structures and serves as the basis for optimization techniques, is crucial in quantum algorithm design.
This paper presents a categorized catalog of quantum circuit ansatzes aimed at supporting quantum algorithm design and implementation. Each ansatz is described with details such as \textit{intent}, \textit{motivation}, \textit{applicability}, \textit{circuit diagram}, \textit{implementation}, \textit{example}, and \textit{see also}. Practical examples are provided to illustrate their application in quantum algorithm design.
The catalog aims to assist quantum algorithm designers by offering insights into the strengths and limitations of different ansatzes, thereby facilitating decision-making for specific tasks.

\end{abstract}


\section{Introduction}
The advent of quantum computing has sparked intensive research and development to harness its potential across various domains, including cryptography~\cite{mosca2018cybersecurity}, optimization~\cite{guerreschi2017practical,farhi2014quantum}, simulation~\cite{childs2018toward}, and machine learning~\cite{dunjko2016quantum,biamonte2017quantum}.
The concept of quantum circuits, sequences of quantum gates that manipulate quantum states to perform computational tasks, is central to quantum computing. Selecting an appropriate quantum circuit ansatz is a critical aspect of quantum algorithm design. 
An ansatz defines the initial structure of a quantum circuit and is a starting point for optimization techniques to find the best parameters representing the desired quantum state~\cite{tilly2022variational}\textit{}.

Quantum circuit ansatzes, in the context of this paper, refer to specific choices of quantum circuits with adjustable parameters. These circuits are designed to prepare or approximate a target quantum state suitable for a particular computational problem. The ansatz is a flexible template that can be fine-tuned through classical optimization algorithms, enabling quantum algorithm designers to tailor circuits to solve specific tasks effectively.

%Importance of a Comprehensive Catalog
There are many types of quantum circuit ansatzes, which pose a significant challenge to researchers and developers. Exploring and comparing different ansatzes becomes complex and time-consuming without a systematic and comprehensive catalog. By providing a centralized resource, we aim to facilitate decision-making for specific tasks in quantum algorithm design.

%Purpose and Structure of the Paper
This paper presents a categorized catalog of quantum circuit ansatzes aimed at supporting quantum algorithm design and implementation. Each ansatz will be thoroughly described, encompassing its \textit{intent}, \textit{motivation}, \textit{applicability}, \textit{circuit diagram}, \textit{implementation}, \textit{example}, and \textit{see also}. We also discuss practical examples to illustrate the application of these ansatzes in quantum algorithm design.

%Significance and Contributions
The catalog of quantum circuit ansatzes proposed in this paper holds significant value for researchers and developers in quantum algorithm design and implementation. By consolidating diverse ansatzes, it will serve as a foundational resource, enabling quantum algorithm designers to explore and adapt various techniques to tackle complex computational problems efficiently. 

The rest of this paper is structured as follows: In Section~\ref{sec:background}, we provide background information, including quantum computing, quantum circuit ansatz, and variational quantum algorithm. Section~\ref{sec:methodology} outlines our method for collecting ansatzes. We present our ansatz catalog in Section~\ref{sec:catalog}. Section~\ref{sec:discussion} is dedicated to discussions. We discuss related work in Section~\ref{sec:related-work}. Finally, we conclude our work in Section~\ref{sec:conclusion}.

\section{Background}
\label{sec:background}

\subsection{Quantum Bit (Qubit)}
A qubit is the basic unit of information in quantum computing. A classical bit is numerical information expressed as 0 or 1, whereas a qubit refers to a state where 0 and 1 exist probabilistically. In other words, a qubit can be a linear combination of two states by exploiting the quantum mechanical phenomenon of superposition. \black{Equation (\ref{qubit_state}) mathematically represents a qubit in a continuum state:}

\begin{equation}
    |\psi\rangle = \alpha|0\rangle + \beta|1\rangle,\ \text{where}\ |\alpha|^2 + |\beta|^2 = 1
    \label{qubit_state}
\end{equation}

\noindent
\black{We cannot definitively determine whether the qubit is in state $|0\rangle$ or $|1\rangle$. Here, $|0\rangle$ and $|1\rangle$ are the ground state represented by column vectors (\ref{eq:vector}), $\alpha$ and $\beta$ denote the amplitudes of the qubit, with measurement probabilities corresponding to $|\alpha|^2$ and $|\beta|^2$. Upon measurement, the probability of observing $|0\rangle$ is $|\alpha|^2$, while the probability of observing $|1\rangle$ is $|\beta|^2$.}

\begin{equation}
  |0\rangle =
     \left(
       \begin{array}{r}
         1 \\
         0 \\
       \end{array}
     \right)\ 
  |1\rangle =
    \left(
      \begin{array}{r}
        0 \\
        1 \\
      \end{array}
    \right)\label{eq:vector}  \\
\end{equation}

\subsection{Quantum Gate and Quantum Circuit}
In classical computing assembly language, operations are typically expressed using logical operations such as 'and,' 'or,' and 'not.' Similarly, we employ quantum gates to execute analogous computations in quantum computing. When applied to N qubits, these gates are generally represented by a $2^N$ x $2^N$ unitary matrix. Quantum gates serve various functions; for instance, the Pauli gate can rotate qubits to arbitrary angles, the Hadamard gate introduces superposition to qubits, and the controlled-not gate establishes controlled connections between qubits. The relationship between a quantum gate and a quantum circuit resembles that of classical logic gates to digital circuits. Quantum circuits are composed of interconnected quantum gates arranged sequentially alongside qubits and measurements.

\subsection{Quantum Circuit Ansatz} 

In quantum computing, an 'Ansatz' refers to a trial wavefunction or trial state used as a starting point for approximations or optimizations~\cite{tilly2022variational}. It is a parameterized quantum state that serves as an educated guess for solving a particular problem, such as finding the ground state of a quantum system. The quantum ansatz is essentially a parameterized quantum circuit, carefully tailored in terms of qubit number, operation count, and operation types to suit the problem. The structure of the ansatz may be informed by prior problem knowledge or physical intuition. Subsequently, these parameters are optimized using classical or quantum methods to approximate the desired quantum state closely. The ansatz is iteratively refined as additional information becomes available. Selecting an appropriate ansatz is crucial in quantum algorithm design, as it significantly impacts the accuracy of the outcomes.

\subsection{Variational Quantum Algorithm}

The Variational Quantum Algorithm (VQA) offers an alternative approach to Quantum Phase Estimation (QPE), addressing hardware limitations~\cite{tilly2022variational}. By selecting a suitable ansatz and optimizing its parameters, VQA approximates the ground state of a given Hamiltonian. Variational Quantum Eigensolver~\cite{peruzzo2014variational}, Quantum Approximate Optimization Algorithm~\cite{farhi2014quantum}, and Quantum Machine Learning~\cite{dunjko2016quantum,biamonte2017quantum} are common types of VQAs.

\subsubsection{Variational Quantum Eigensolver (VQE)}

VQE targets ground and excited state energies of quantum systems. It employs parameterized quantum circuits and classical optimization algorithms to minimize energy. The key concept is using parameterized quantum circuits to approximate the lowest energy eigenstates and corresponding eigenvalues of a given Hamiltonian. By adjusting circuit parameters, VQE seeks the quantum state, minimizing the Hamiltonian's expected value. This property finds applications in solving quantum chemistry problems. However, VQE's applicability is limited by the exponential growth of the Hamiltonian dimension, often surpassing current quantum computing capabilities.

\subsubsection{Quantum Approximate Optimization Algorithm (QAOA)}

QAOA
%\footnote{To avoid confusion, we will consistently use the abbreviation QAOA throughout this paper to refer to the Quantum Approximate Optimization Algorithm, also known as the Quantum Alternating Operator Ansatze.} 
is a variational quantum algorithm. At its core, the QAOA involves constructing a parameterized quantum circuit that encodes the optimization problem into the parameter space of a quantum state. This approach harnesses the parallelism and quantum nature of quantum states, potentially achieving exponential speedups for specific problems. However, the QAOA faces challenges such as selecting the appropriate quantum circuit structure and parameters, as well as obtaining reliable optimization results within quantum hardware's noise and error constraints.

\subsubsection{Quantum Machine Learning (QML)}

QML integrates principles from quantum computing and machine learning to leverage quantum properties such as superposition and entanglement. Unlike classical computers, which require $log(N)$ qubits to store N-dimensional data, quantum computers offer the potential for exponential speedup over classical counterparts in specific tasks. However, QML still faces challenges, such as limitations in the current scale of quantum computers and issues related to noise, decoherence, and the barren plateau problem.

However, the challenges mentioned for VQE, QAOA, and QML are not insurmountable. Numerous studies have effectively addressed these challenges by developing appropriate approaches to specific issues. Therefore, the choice of ansatz structure plays a crucial role in the VQAs.

\section{Method for Collecting Ansatzes}
\label{sec:methodology}

\black{This section presents the methodology adopted for developing the catalog of quantum circuit ansatzes. We systematically collected and analyzed the literature on quantum circuit ansatzes to create a comprehensive catalog.
}

\black{First, we used keywords such as "quantum ansatz," "quantum circuit design," and "variational quantum algorithms" to search multiple databases, including arXiv, Google Scholar, and IEEE Xplore. We then conducted an initial screening based on titles and abstracts, followed by a thorough review of potentially relevant papers, focusing on extracting information related explicitly to quantum ansatzes.
After the initial screening, we meticulously reviewed over 60 papers, extracting key information on the types of ansatzes, their application domains, advantages, and limitations. This information was then categorized and summarized according to the kind of problems they addressed.
}

%This section presents the methodology adopted for developing the catalog of quantum circuit ansatzes. The process involves data collection and the creation of an organized framework to categorize and describe the ansatzes.

%The process began with a comprehensive search for "quantum ansatz" across various databases, including arXiv and Google Scholar. Over 60 papers and websites were meticulously reviewed, focusing on extracting information relevant to quantum circuit ansatzes.

In the catalog, each ansatz is described with \textit{intent}, \textit{motivation}, \textit{applicability}, \textit{circuit diagram}, \textit{implementation}, \textit{example}, and \textit{see also}. The detailed description of each item within an ansatz is provided as follows:

\begin{itemize}[leftmargin=1.5em]

\item \textbf{Intent}: Clarifies the purpose of an ansatz, addressing the design challenges it aims to resolve and to provide insights into specific design considerations.

\item \textbf{Motivation}: Outlines the specific design challenges or issues addressed by the ansatz, along with its structural components tailored to tackle these challenges.

\item \textbf{Applicability}: Explains the practical application scenarios for the ansatz and provides guidance on its usage.

\item \textbf{Circuit Diagram}: Depicts the structure and conceptual framework of the ansatz using the quantum circuit model.

\item \textbf{Implementation}: Outlines potential pitfalls and considerations when implementing the ansatz, offering insights to navigate the process effectively.

\item \textbf{Example}: Showcases real-world scenarios where the ansatz is employed to address specific problems. While the aim was to include two distinct examples for each ansatz, some instances only have one.

\item \textbf{See Also}: Lists closely related ansatzes regarding the intent, explains their relationship, and highlights important distinctions between them.
\end{itemize}

\begin{comment}
    \item \textbf{Intent}: Serves to elucidate the purpose of an ansatz, clarifying its intention and addressing the design challenges it aims to resolve. It explains the rationale behind creation and provides insights into specific design considerations it addresses.
    
    \item \textbf{Motivation}: Outline the specific design challenges or issues in crafting an ansatz and the structural components and defining characteristics tailored to address these challenges.

    \item \textbf{Applicability}: Explains the situations where ansatz can be applied effectively and provides guidance on application scenarios.

    \item \textbf{Circuit Diagram}: Depicting ansatz's structure and conceptual framework and illustrating ansatz using the quantum circuit model. 

    \item \textbf{Implementation}: Outline the potential pitfalls and considerations when implementing ansatz, providing insights to navigate the process effectively.
   
    \item \textbf{Example}: This section showcases when ansatz is employed to address specific problems in real-world scenarios. While we tried to include two distinct examples for an ansatz, some instances were found to have only one. This section provides an example of applying ansatz in real-world scenarios. 

    \item \textbf{See Also}: We list ansatz as closely related in intent to current ansatz. Additionally, explaining their relationship and highlight important distinctions between them.
\end{comment}

\section{A Catalog of Quantum Circuit Ansatzes}
\label{sec:catalog}

In this section, we classify quantum circuit ansatzes into distinct categories based on VQE, QAOA, and QML discussed in Section~\ref{sec:background}. Each category represents a collection of ansatzes that share fundamental characteristics, enabling readers to gain deeper insights into the diverse landscape of quantum circuit ansatzes and their applications in quantum algorithm design.

Table~\ref{table:ansatz} presents the ansatzes collected in our catalog. We propose a categorization method to summarize existing ansatzes, defining four sub-categories: the problems addressed by the ansatz in VQAs, the names of the ansatzes, their extended versions (referenced in \textit{see also}), and descriptions of each ansatz. In the following, we select ten ansatzes; each will be given a detailed description in terms of its \textit{intent}, \textit{motivation}, \textit{applicability}, \textit{circuit diagram}, \textit{implementation}, \textit{example}, and \textit{see also}.  

%%%%%%%%%%%%%%%%%%%%%%%%%%% TABLE %%%%%%%%%%%%%%%%%%%%%%%%%%%%%%%%%%%%%%%%%%%%%%%%%%%%
% \begin{table}
\begin{table*}[ht]
% 表设置于文本中央
\centering
\caption{A Catalog of Quantum Circuit Ansatzes}
\label{table:ansatz}
\renewcommand\arraystretch{1.0}
%\footnotesize
\scriptsize
\begin{tabular}{|p{0.7cm}|p{3.8cm}|p{5.5cm}|p{6.3cm}|} %l(left)居左显示 r(right)居右显示 c居中显示
\hline 
{\bf VQA} \hspace{0.3mm}  & Ansatz & Extensions of Each Ansatz & Description \\
\hline 
\multirow{10}{*}{\tabincell{l}{{\bf VQE} }} 
                        & \multirow{3}{*}{\makecell[l]{Unitary Coupled Cluster (UCC) \\Ansatz~\cite{shen2017quantum} [Sec.\ref{UCC}]}} 
                                                   &Unitary Vibrational Coupled Cluster (UVCC) Ansatz~\cite{mcardle2019digital}    & It is the scalable scheme for generating the parameterized states required for variational methods. This heuristic ansatz is widely used in quantum chemistry problems \\\cline{3-3}
                        &                          &Generalized UCC (UCCG) $\&$ Unitary Coupled Cluster with Single and Double excitations (UCCSD)~\cite{romero2018strategies}  & \\ \cline{3-3} 
                        &                          & k-qubit Universal Parallel Circuit with Coupled Cluster Generalized Single  
                                                    and Double excitations (k-UpCCGSD)~\cite{lee2018generalized}   & \\\cline{3-3}
                        &                          & Orbital Optimized UCC (OO-UCC)~\cite{mizukami2020orbital}  & \\ \cline{3-3}
                        &                          & Unitary Cluster-Jastrow Ansatz~\cite{matsuzawa2020jastrow} & \\ \cline{3-3}
                        &                          & Low Depth Circuit Ansatz (LDCA)~\cite{dallaire2019low} &  \\ \cline{2-4}
                         & \makecell[l]{Hardware-Efficient Ansatz (HEA)\\~\cite{kandala2017hardware} [Sec.\ref{HEA}]} &Qubit Coupled-Cluster Ansatz (QCC)~\cite{ryabinkin2018qubit}, iterative QCC (iQCC)~\cite{ryabinkin2020iterative} & It customizes the initialization state for QVE problems to specific quantum devices.\\ \cline{2-4}

                        & \multirow{2}{*}{\makecell[l]{Adaptive Derivative Assembled \\Pseudo-Trotter (ADAPT VQE) \\Ansatz~\cite{grimsley2019adaptive} [Sec.\ref{ADAPT VQE}]}} & qubit ADAPT VQE~\cite{tang2021qubit} & An adaptive ansatz aimed at incrementally constructing a parametric representation of quantum states, reducing circuit depth to achieve higher accuracy.\\\cline{3-3}
                        & &\makecell[l]{Qubit-excitation-based adaptive VQE \\(QEB ADAPT VQE)~\cite{yordanov2021qubit}} & \\\cline{3-3}
                        & &ClusterVQE~\cite{zhang2022variational} & \\ \cline{2-4}
                        & Symmetry-Preserving Ansatz (SPA)~\cite{barkoutsos2018quantum} [Sec.\ref{Symmetry-Preserving}] &Efficient Symmetry-Preserving Ansatz (ESPA)~\cite{gard2020efficient} & An ansatz constructing a parameterized representation of quantum states, aimed at ensuring that the generated quantum states remain invariant under specific symmetry operations. \\\cline{2-4}       
                        & Quantum Circuit Matrix Product State (QCMPS)  Ansatz~\cite{fan2023quantum} \hspace{0.3mm} & & An ansatz to calculate the ground state energies of large molecular systems with very few qubits and quantum gates \\\cline{2-4}
                        & Isometric Tensor-Network States (isoTNS)~\cite{soejima2020isometric} \hspace{0.3mm} & & An ansatz to calculate the ground state energies of large molecular systems with very few qubits and quantum gates. \\\cline{2-4}    
                        & Pairwise Perturbative Ansatz (PAPA)~\cite{govia2020bootstrapping} \hspace{0.3mm} & & An ansatz to limit the TNS and allow for a highly efficient network contraction. \\    
\hline
\multirow{4}{*}{\tabincell{l}{{\bf QAOA} }} & \multirow{2}{*}{\tabincell{l}{Quantum Alternating Operator \\Ansatz (QAOA)~\cite{farhi2014quantum} [Sec.\ref{QAOA}]}} & Grover-Mixer QAOA (GM-QAOA)~\cite{bartschi2020grover} \hspace{0.3mm}  & It customizes an alternating structure in its ansatz to address Quantum Approximate Optimization Algorithm obtaining approximate solutions for combinatorial optimization problems. \\\cline{3-3}

                         &                          & Augmented QAOA (QAOA+)~\cite{chalupnik2022augmenting}   &\\\cline{2-4}

                         & \multirow{2}{*}{\makecell[l]{Hamiltonian Variational Ansatz \\(HVA)~\cite{wecker2015progress} \hspace{0.3mm} [Sec.\ref{HVA}]}}              &Symmetry Breaking HVA~\cite{babbush2018low} & A circuit design approach based on hierarchical structure and adjustable parameters aimed at more effectively managing and controlling the complexity of quantum circuits to tackle complex quantum computing problems. \\\cline{3-3}
                         &                          &Variational Extended Hamiltonian Ansatz (VMFHA)~\cite{vogt2020preparing}  &\\ \cline{3-3}
                         &                          &quantum-optimal-control-inspired ansatz (QOCA)~\cite{choquette2021quantum} & \\ \cline{3-3}
            
                         &                          & Fourier-transform HVA~\cite{vogt2020preparing}   &\\
                                               
\hline

\multirow{5}{*}{{\bf QML} } & \multirow{2}{*}{\makecell[l]{Quantum Circuit Embedding \\(QCE)\cite{lloyd2020quantum}  [Sec.\ref{QCE}]}} & \makecell[l]{Fully Parameterized Quantum \\Circuit Embedding ({\tiny F}QCE)~\cite{ma2019variational}} & An ansatz for encoding conventional data into quantum states. \\\cline{3-3}
                                        & & \black{Quantum Embedding Kernels (QEKs)}~\cite{hubregtsen2022training}  &\\\cline{2-4}
                                        & Multiscale Entanglement Renormalization Ansatz (MERA)~\cite{vidal2008class} \hspace{0.3mm} [Sec.\ref{MERA}] & & An ansatz for quantum many-body states on a D-dimensional lattice precisely and efficiently calculate the local observables' expectation \\\cline{2-4}              
                                        & Quanvolutional Neural Network (QNN)~\cite{henderson2020quanvolutional} \hspace{0.3mm} [Sec.\ref{QNN}] & & A hybrid classical-quantum algorithm leveraging some non-linear quantum circuit transformations for CNN.\\\cline{2-4}              
                                        & Quantum Convolutional Neural Network (QCNN)~\cite{cong2019quantum}  [Sec.\ref{QCNN}] & & A quantum convolutional neural network inspired by an inversed MERA circuit to enable efficient machine learning on quantum devices. \\ 
\hline

\end{tabular}

%Four sub-categories, including problems addressed, ansatz names, their extended versions (referenced in \textit{see also}), and descriptions of each ansatz.

\end{table*}

%UCC
\subsection{Unitary Coupled Cluster (UCC) Ansatz}
\label{UCC}

\begin{itemize}[leftmargin=1.5em]

\item \textbf{Intent}:
Developing an efficient computational scheme for UCC Ansatz~\cite{shen2017quantum} to obtain molecular ground states has been a longstanding challenge. The UCC approach offers a generic and scalable scheme for generating parameterized states required for variational methods, and it can be implemented efficiently in quantum devices with trapped ions.

\item \textbf{Motivation}: 
The inefficiency of unitary modification in classical computers limits the application of classical coupled-cluster ansatz. UCC aims to express the wave function of a quantum system, based on exciting some reference state $\ket{\phi_0}$ as $e^{T(\theta)-T(\theta)^\dag}\ket{\phi_0}$, where $T = \sum_K T_K$ represents the cluster operator and $T_k$ are excitation operators~\cite{cerezo2021variational}.
The UCC ansatz is implemented on a quantum computer by decomposing the unitary into a quantum gate. After efficiently generating trial states using the UCC ansatz in a quantum system mapping the classical ground state of the molecule of interest, the average energy of the trial states is measured. The UCC parameters are then adjusted based on a classical feedback algorithm. The molecular ground state can then be obtained by repeating the quantum process of preparation and measurement until the target Hamiltonian variational minimum is found.

\item \textbf{Applicability}:
UCC analysis is applicable for efficiently obtaining the ground state of a molecule, surpassing the efficiency of classical computer-based UCC analysis.

\item \textbf{Circuit Diagram}: Please refer to Figure~\ref{fig:UCC}.

\begin{figure}[h]
  \centering
  \resizebox{0.47\textwidth}{!}{
  \begin{quantikz}[row sep=0.5cm,column sep=0.3cm,wire types={q,q}]
    %qubit 0
    & \gate{H} &\ctrl{1}  & &\ctrl{1} &\gate{H} &\gate{R_X(\pi/2)} &\ctrl{1} & &\ctrl{1}  &\gate{R_X(-\pi/2)}&\\ 
    %qubti 1
    & \gate{R_X(\pi/2)} &\targ{} &\gate{R_Z(\theta)} &\targ{} &\gate{R_X(-\pi/2)} &\gate{H} &\targ{} &\gate{R_Z(\theta)} &\targ{} &\gate{H} &
  \end{quantikz}
}
  \caption[Unitary coupled cluster (UCC) ansatz]{Unitary Coupled Cluster (UCC) ansatz~\cite{anand2022quantum}}
  \label{fig:UCC}
\end{figure}
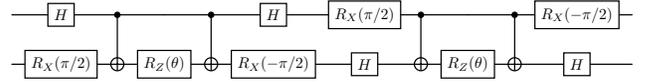

\item \textbf{Implementation}:
Since the UCC method is not hardware efficient, effectively relocating qubits together during implementation using SWAP gates can reduce the overhead caused by a large number of CNOT gates.

\item \textbf{Example}:
Shen {\it et al.}~\cite{shen2017quantum} used the UCC approach to simulate the electronic structure of molecular ions (HeH+ ), searching for the ground state energy curve, studying the energies of the excited states, and simulating bond dissociation in a non-perturbative manner. Liu {\it et al.}~\cite{liu2020simulating} use UCC to simulate periodic systems on quantum computers.

\item \textbf{See Also}:
The UCC approach can simulate VQE problems and effectively reduce circuit depth. 
%UCCSD
Depending on the different ways of composing the single-particle and two-particle excitation operators, different UCC methods are composed. For example, the UCCSD method~\cite{romero2018strategies} considers only the single-particle and two-particle excitations to approximate the ground-state wavefunction of the molecule, which can effectively reduce the depth of the UCC circuit without significant loss of accuracy.
The k-UpCCGSD method~\cite{lee2018generalized} effectively controls the number of excitations by introducing the parameter k, which restricts the excitation count for each electron pair. Additionally, the method enhances computational accuracy and stability by incorporating mathematical concepts such as symmetry and group theory.
\end{itemize}

\subsection{Hardware-Efficient Ansatz (HEA)}
\label{HEA}
\begin{itemize}[leftmargin=1.5em]

\item \textbf{Intent}:
The Hardware-Efficient Ansatz (HEA)~\cite{kandala2017hardware} enables the parameterization of trial states with quantum gates and customization to available physical devices. Unlike the UCC ansatz~\cite{shen2017quantum}, which relies on high-fidelity quantum gates approximating tuned unitary operators based on a theoretical Ansatz, HEA does not require precise implementation of specific two-qubit gates.

\item \textbf{Motivation}:
The HEA method tailors the initialization state for VQE problems to specific quantum devices. The initial motivation is to transform arbitrary unitary operations into combinations of gate sequences that can be easily implemented on current quantum devices, thereby reducing the circuit depth for optimization. Therefore, the HEA method typically follows the same logic, composed of quantum blocks as shown in Figure~\ref{fig:hardware-efficient ansatz}. Each quantum block contains parameterized single-qubit rotations and ladders of entanglers, which are adapted based on the gate set on the quantum device and the target state.

\item \textbf{Applicability}:
HEA can optimize the Hamiltonian problem using no more than six qubits and over 100 Pauli terms.

\item \textbf{Circuit Diagram}:
Please refer to Figure~\ref{fig:hardware-efficient ansatz}.

\begin{figure}[h]
  \centering
  \resizebox{0.47\textwidth}{!}{
  \begin{quantikz}[row sep=0.5cm,column sep=0.4cm,wire types={q,q,q,q}]%
    %qubit 0
    && \gate{R_X(\theta_1)}\gategroup[4,steps=3,style={dashed, inner xsep=6pt}]{Single-qubit rotations} &\gate{R_Z(\theta_2)} &\gate{R_X(\theta_3)} && &\ctrl{1}\gategroup[4,steps=4,style={dashed, inner xsep=6pt}]{Entangler} &&  &\gate{R_Y(\theta_{16})} &&\\
    %qubit 1
    && \gate{R_X(\theta_4)} &\gate{R_Z(\theta_5)} &\gate{R_X(\theta_6)} &&&\gate{R_Y(\theta_{13})}  &\ctrl{1} &&&&\\ 
    %qubit 2
    && \gate{R_X(\theta_7)} &\gate{R_Z(\theta_8)} &\gate{R_X(\theta_9)} &&& &\gate{R_Y(\theta_{14})} &\ctrl{1} &&&\\
    %qubit 3
    && \gate{R_X(\theta_{10})} &\gate{R_Z(\theta_{11})} &\gate{R_X(\theta_{12})} &&&& &\gate{R_Y(\theta_{15})} &\ctrl{-3}&&
\end{quantikz}
}
  \caption{Hardware-Efficient Ansatz (HEA)~\cite{kardashin2020certified}}
  \label{fig:hardware-efficient ansatz}
\end{figure}
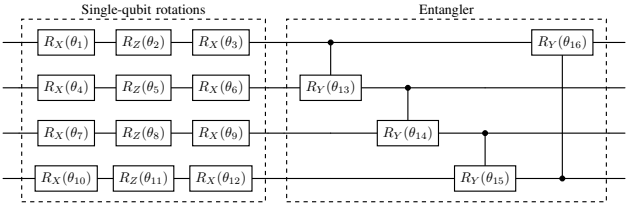

\item \textbf{Implementation}: 
When implementing HEA, it is crucial to ensure compatibility with the quantum hardware, design the approach based on the characteristics of the current quantum device, and minimize errors introduced by the hardware.

\item \textbf{Example}:
Kandala {\it et al.}~\cite{kandala2017hardware} optimized the molecular energies of $H_2$, $LiH$, and $BeH_2$ using HEA on a superconducting quantum processor.

\item \textbf{See Also}:
HEA is a specific methodology for designing parametric quantum circuits that consider the limitations of current quantum hardware. Since HEA optimizes the hardware feasibility of quantum circuits, it is ideally suited for implementation on real quantum computers and solving VQE problems. The Qubit Coupled-Cluster (QCC) ansatz~\cite{ryabinkin2018qubit} directly implements the coupled cluster approach (see Section~\ref{UCC}) in the qubit space. The QCC method is restricted to a two-qubit entanglement gate with a factorization technique to allow arbitrary multi-qubit entanglers on limited qubit hardware, enabling highly efficient use of quantum resources in terms of coupled cluster operators. The iterative Qubit Coupled Cluster method (iQCC)~\cite{ryabinkin2020iterative} enhances QCC by iteratively incorporating operators into the Hamiltonian, reducing the ansatz operators.
\end{itemize}

%Adaptive VQE
\subsection{Adaptive Derivative Assembled Pseudo-Trotter (ADAPT VQE) Ansatz}
\label{ADAPT VQE}
\begin{itemize}[leftmargin=1.5em]

\item \textbf{Intent}:
While UCCSD truncation retains only single and double excitation operators in the ansatz, it still entails numerous qubit entanglement gates. Consequently, this leads to elevated circuit depth and necessitates a substantial number of optimization parameters. ADAPT VQE~\cite{grimsley2019adaptive} addresses these by automatically adapting to the computing resources.

\item \textbf{Motivation}: 
The idea of ADAPT VQE is to progressively build the ansatz by sequentially adding operators that contribute the most to lowering the VQE energy to the ground state. The ADAPT VQE ansatz employs a dynamic approach rather than a fixed quantum circuit to achieve adaptability. This method selects operators from a predefined operator pool to construct the ansatz dynamically. During the dynamic construction process, the ansatz is iteratively expanded, with each iteration selecting operators from the operator pool with the greatest energy impact. In other words, this algorithm reduces circuit depth and the number of parameters by increasing the number of measurements.

\item \textbf{Applicability}:
Simulating with NISQ devices to achieve higher precision or efficiency within limited computational resources.

\item \textbf{Circuit Diagram}:
Please refer to Figure~\ref{fig:ADAPT}.

\begin{figure}[h]
  \centering
  \includegraphics[width=0.7\linewidth]{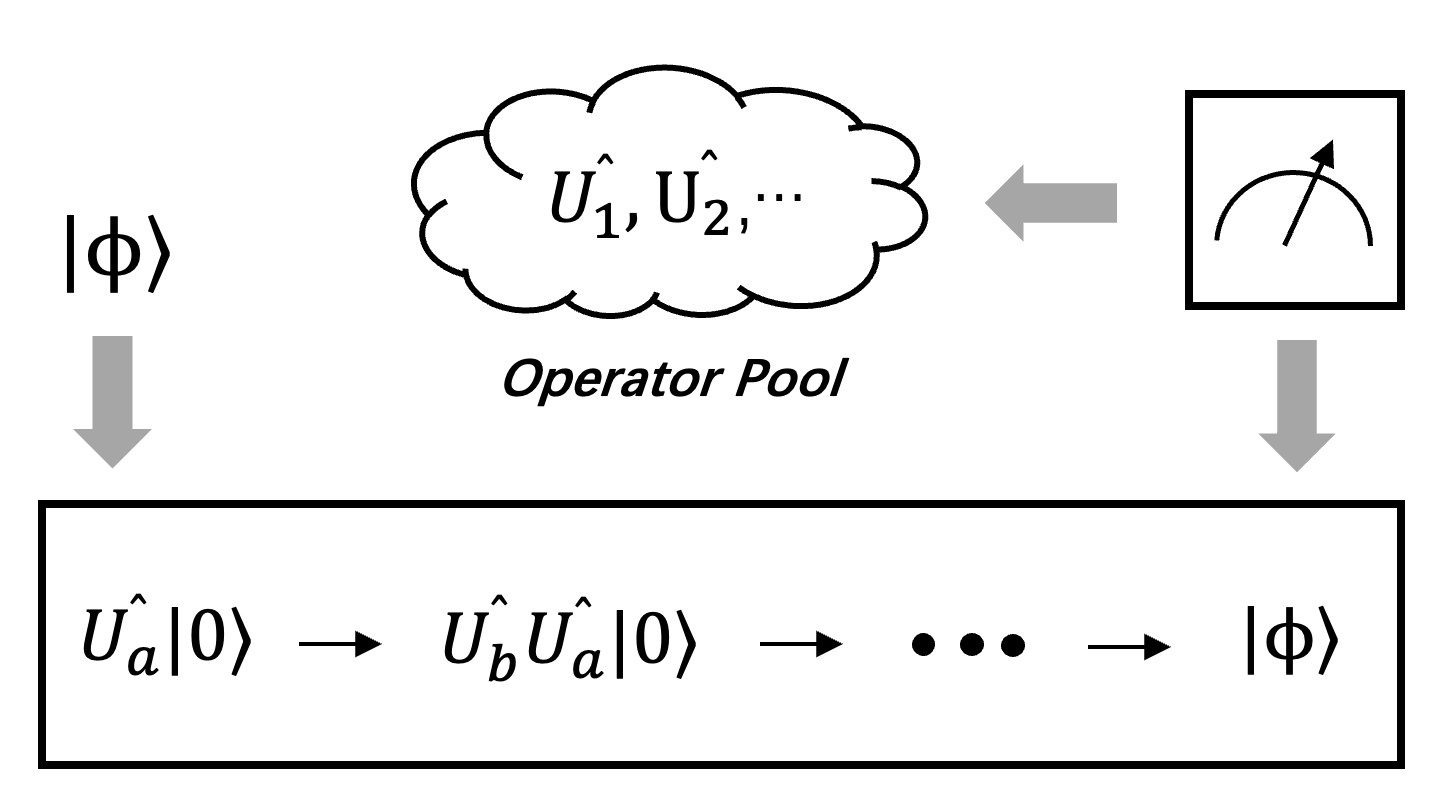}
  \caption[Hamiltonian Variational Ansatz (HVA) process]{Hamiltonian Variational Ansatz (HVA)~\cite{tang2021qubit}}
  \label{fig:ADAPT}
\end{figure}

\item \textbf{Implementation}:
The order of operators is crucial for the ADAPT VQE ansatz; an appropriate order can lead to faster convergence with shallower quantum circuits. Hence, each parameter is variable and differs across different iterations.

\item \textbf{Example}:
Grimsley {\it et al.}~\cite{grimsley2019adaptive} demonstrated that, even when the operator pool contains only single and double excitations, ADAPT VQE outperforms UCCSD in terms of both parameter count and accuracy. Liu {\it et al.}~\cite{liu2020simulating} use ADAPT VQE to simulate periodic system ground state on a quantum computer.

\item \textbf{See Also}:
The ADAPT VQE method improves upon UCCSD by dynamically constructing quantum circuits to reduce circuit depth and achieve higher precision. However, the resulting ansatz may still exceed the capabilities of NISQ devices. Therefore, Tang {\it et al.}~\cite{tang2021qubit} proposed the qubit-ADAPT-VQE method, which utilizes a qubit pool to generate shallower ansatz circuits with fewer CNOT gates at the cost of introducing more variational parameters.
\black{Yordanov {\it et al.}~\cite{yordanov2021qubit} proposed an ansatz circuit maintains the same parameters as ADAPT QVE and fewer CNOT fate than both ADAPT VQE and qubit-ADAPT-VQE.}
\end{itemize}

\subsection{Symmetry-Preserving Ansatz (SPA)}
\label{Symmetry-Preserving}
\begin{itemize}[leftmargin=1.5em]

\item \textbf{Intent}:
The performance of VQEs critically depends on the form of the variational ansatz. A suitable ansatz translates to relatively shallow circuits and involves few classical optimization parameters. Symmetries play a central role in preparing multi-qubit trial states on quantum devices, affecting the efficiency of the VQE.

\item \textbf{Motivation}: 
Symmetry-preserving\cite{barkoutsos2018quantum} includes the minimum number of parameters necessary to cover the symmetry subspace completely, avoids all states outside this subspace, and ensures that the true ground state is contained within the state space covered by the circuit.

Symmetry preservation is typically achieved through exchange-type gates, aiming to induce relative phase changes between superposition states like \ket{01} and \ket{10} while keeping states like \ket{00} and \ket{11} unchanged. These exchange-type gates comprise three CNOT gates and two rotation gates.

\item \textbf{Applicability}:
Constraining the size of the Hilbert space can accelerate optimization, reduce the risk of barren plateaus, and help avoid local inaccuracies. However, the large span of the Hilbert space can lead to lower efficiency of ansatz based on symmetry preservation.

\item \textbf{Circuit Diagram}: 
Please refer to Figure~\ref{fig:symmetry-preserving}.

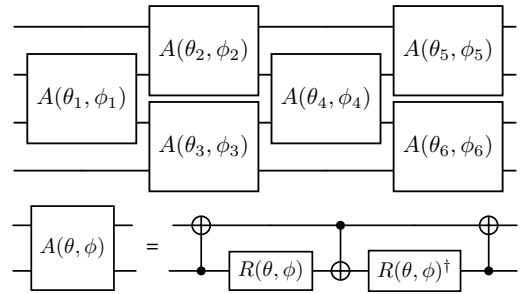
\begin{figure}[h]
  \centering
 \resizebox{0.8\linewidth}{!}{
  \begin{quantikz}[row sep=0.1cm,column sep=0.2cm,wire types={q,q,q,q}]
    %qubit 0
    & & \gate[2]{A(\theta_2,\phi_2)} & &\gate[2]{A(\theta_5,\phi_5)}&\\ 
    %qubit 1
    & \gate[2]{A(\theta_1,\phi_1)} & & \gate[2]{A(\theta_4,\phi_4)}&&\\
    %qubti 2
    & &\gate[2]{A(\theta_3,\phi_3)} & &\gate[2]{A(\theta_6,\phi_6)}&\\
    %qubti 3
    & & & & &
  \end{quantikz}
}
  \resizebox{0.8\linewidth}{!}{
  \begin{quantikz}[row sep=0.1cm,column sep=0.3cm,wire types={q,q}]
    %qubit 0
    & \gate[2]{A(\theta,\phi)} & \midstick[2,brackets=none]{=} & \targ{} & &\ctrl{1} & &\targ{} &\\ 
    %qubti 3
    &  & &  \ctrl{-1} &\gate{R(\theta,\phi)} &\targ{} &\gate{R(\theta,\phi)^{\dag}} &\ctrl{-1}&
  \end{quantikz}
}
  \caption[Symmetry-preserving]{Symmetry-Preserving ansatz (SPA)~\cite{endo2020calculation}}
  \label{fig:symmetry-preserving}
\end{figure}

\item \textbf{Implementation}:
To preserve symmetry, a modification can be made to the cost function, as explained in the paper by Tilly \textit{et al.}~\cite{tilly2022variational}.

\item \textbf{Example}:
Gard {\it et al.}~\cite{gard2020efficient} tested symmetry-preserving in quantum simulations of $H_2$ and $LiH$ molecules and found that it outperformed standard state preparation methods in both accuracy and circuit depth. Endo {\it et al.}~\cite{endo2020calculation} used symmetry-preserving to preserve the total number of particles in the system when calculating the five Fermi-Hubbard energy eigenstates.

\item \textbf{See Also}:
Symmetry-preserving ansatz reduces the complexity of the classical optimizer in the VQE algorithm. Gard {\it et al.}~\cite{gard2020efficient} proposed Efficient Symmetry-Preserving (ESP) ansatz to address a broader range of symmetry issues, including particle number, time reversal, total spin, and spin magnetization.
\end{itemize}

\subsection{Quantum Alternating Operator Ansatz (QAOA) \protect\footnote{\black{Since the Quantum Alternating Operator Ansatz is an extension of the Quantum Approximate Optimization Algorithm~\cite{farhi2014quantum}, allowing alternation between more general families of operators, we use QAOA in this paper to represent both the Quantum Alternating Operator Ansatz and the Quantum Approximate Optimization Algorithm interchangeably.}}} 
\label{QAOA}
\begin{itemize}[leftmargin=1.5em]

\item \textbf{Intent}:
Farhi {\it et al.}~\cite{farhi2014quantum} presented a quantum approximate optimization algorithm that outperformed the current best approximation bound for efficient classical algorithms for the E3LIN2 problem.
This approach is especially beneficial for optimization problems with strict constraints that must always be met and flexible constraints one aims to minimize. It enables quicker experimental exploration of different approximate and exact optimization and sampling problems.

\item \textbf{Motivation}:
QAOA exploits the potential of quantum computers to search for solutions to combinatorial optimization problems through parametric transformations of quantum states.
Figure \ref{fig:QAOA} illustrates an example decomposition of a QAOA parameterized ansatz. The cost operators, $U_P(\gamma)$, depend on the objective function to be optimized, while the mixer operator, $U_M(\beta)$, depends on the domain and its structure. The parameters $\beta$ and $\gamma$ are optimized using a classical optimizer to get the ansatz as close to the target state.

The mixer operators corresponding to the one-parameter family do not result from time evolution under the fixed-mixing Hamiltonian $H_M$. Due to the different ordering of the partial mixers, several non-equivalent mixing operators exist with varying implementation costs. A more general one-parameter family of unitaries can achieve more efficient implementation while preserving feasible subspaces. Given a domain, its configuration space encoding, phase separator, and mixer, many ways exist to compile the phase separator and mixer into a circuit that works on qubits. 

\item \textbf{Applicability}: 
This approach is beneficial when optimizing for a solution that must satisfy hard and soft constraints when the feasible subspace is smaller than the total space.

\item \textbf{Circuit Diagram}: 
Please refer to Figure~\ref{fig:QAOA}.

\begin{figure}[h]

\centering
\resizebox{0.7\linewidth}{!}{
\begin{quantikz}[row sep=0.3cm,column sep=0.3cm,wire types={q,q,q}]
    %qubit 0
    \lstick{\ket{s_1}}& \gate[3]{U_{P(\gamma_1)}} &\gate[3]{U_M(\beta_1)} &\gate[3]{U_{P(\gamma_2)}} &\midstick[3,brackets=none]{...} &\gate[3]{U_{P(\gamma_P)}} &\gate[3]{U_{M(\beta_P)}} &\meter{}\\ 
    %qubit 1
    \vdots & &&&&&& \vdots\\
    %qubti 2
    \lstick{\ket{s_n}}& &&&&& &\meter{}
\end{quantikz}
}

\centering
\resizebox{0.6\linewidth}{!}{
\begin{quantikz}[row sep=0.3cm,column sep=0.3cm,wire types={q,q,q}]
    %qubit 0
    &\gate[4][2.4cm]{U_{M(\beta_1)}}  &\midstick[4,brackets=none]{=} &\gate[2]{U_{M,1(\beta_1)}} && \gate[2]{U_{M,K(\beta_1)}}&\\ 
    %qubit 1
    && & &\gate[2]{U_{M,j(\beta_1)}} &&\\
    %qubti 2
    && &\gate[2]{U_{M,2(\beta_1)}} &&\gate[2]{U_{M,K{_+{_1(\beta_1)}}}}&\\
    %qubti 3
    &&& &&&
\end{quantikz}
}
  \caption[Quantum Alternating Operator Ansatz (QAOA)]{Quantum Alternating Operator Ansatz (QAOA)~\cite{hadfield2019quantum}}
  \label{fig:QAOA}
   \vspace*{-0.3cm}
\end{figure}
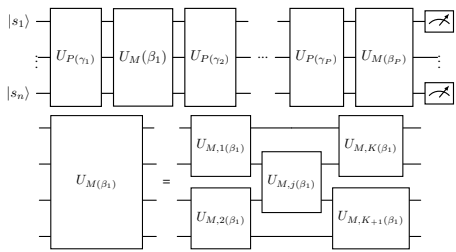

\item \textbf{Implementation}:
Phase and mixing operators frequently use ancilla qubits to simplify computation and compilation into 1- and 2-qubit gates.

\item \textbf{Example}:
Hadfield {\it et al.
}~\cite{hadfield2019quantum} proposed a QAOA mapping method for various optimization problems. They implemented and evaluated the method for portfolio rebalancing.

\item \textbf{See Also}:
\black{The Quantum Approximate Optimization Algorithm (QAOA) is the basis of the Quantum Alternating Operator Ansatz (QAOA), which utilizes an alternating structure in its ansatz during implementation.} It uses special strategies in quantum circuit design and parameter tuning for specific types of combinatorial optimization problems.
GM-QAOA~\cite{bartschi2020grover} utilizes Grover's selective phase shift operator to obtain a simple mixing unitary that provides good
transition properties in the feasible solution space
QAOA+~\cite{chalupnik2022augmenting} improves the approximation ratio through an additional multiparameter problem-independent layer.
\end{itemize}

\subsection{Hamiltonian Variational Ansatz (HVA)}
\label{HVA}
\begin{itemize}[leftmargin=1.5em]

\item \textbf{Intent}:
Conventional quantum algorithms typically require the design of complex quantum circuits to handle problems with highly entangled properties. The Hamiltonian Variational Ansatz (HVA)~\cite{wecker2015progress} introduces a hierarchical structure and adjustable parameters to more effectively manage and control the complexity of quantum circuits, making it easier to solve complex quantum computing problems. By using short quantum circuits to prepare quantum states, the fidelity of quantum gates can be ensured if the circuit is short enough, allowing execution without the need for quantum error correction. This is one of the most promising applications for current small-scale quantum computers. HVA applies this concept to variational approaches, addressing a given Hamiltonian problem with a modest gate depth and a very limited number of variational parameters compared to the system size.

\item \textbf{Motivation}:
The inspiration for HEA is drawn from both QAOA~\cite{hadfield2019quantum} and adiabatic quantum computation~\cite{farhi2000quantum}. To achieve the alternative layer concept in QAOA, HEA employs Trotterization during the adiabatic state preparation process, decomposing the Hamiltonian into Hermitian operators. Each Trotter step corresponds to a variational ansatz, and the variational ansatz constructed through layer-by-layer training facilitates optimization.

\item \textbf{Applicability}:
HEA plays a critical role in VQE by mitigating or avoiding barren plateau issues and addressing the overparametrization phenomenon. It also exhibits scalability with system size, making it suitable for ground state estimation for complex Hamiltonians.

\item \textbf{Circuit Diagram}:
Please refer to Figure~\ref{fig:HVA}.

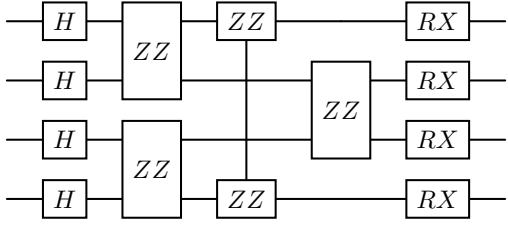
\begin{figure}[t]
  \centering
    \resizebox{0.8\linewidth}{!}{
    \begin{quantikz}[row sep=0.3cm,wire types={q,q,q,q}]%
    %qubit 0
    &\gate{H} &\gate[2]{ZZ} &\gate{ZZ} &             &\gate{RX}&\\
    %qubit 1
    &\gate{H} &             &          &\gate[2]{ZZ}   &\gate{RX}&\\ 
    %qubit 2
    &\gate{H} &\gate[2]{ZZ} &          &        &\gate{RX}&\\
    %qubit 3
    &\gate{H} &            &\gate{ZZ}\wire[u][3]{q} &  &\gate{RX}&
    \end{quantikz}
    }
  \caption[Hamiltonian Variational Ansatz (HVA)]{Hamiltonian Variational Ansatz (HVA) for TFIM with p = 1~\cite{wiersema2020exploring}}
  \label{fig:HVA}
    \vspace*{-0.3cm}
\end{figure}

\item \textbf{Example}:
Wecker {\it et al.}~\cite{wecker2015progress} found that HEA converges faster than the previously proposed UCC method.
Wiersema {\it et al.}~\cite{wiersema2020exploring} shows that HEA has a solid resilience for barren plateaus, as evidenced by the fact that even when overparameterized, HEA can maintain an optimization landscape that is almost free of traps.

\item \textbf{See Also}:
The HVA is based on the concept of alternative layers in the QAOA. It aims to effectively manage and control the complexity of quantum circuits by introducing hierarchical structures and adjustable parameters to tackle complex quantum computing problems. In contrast, QAOA is primarily used for solving combinatorial optimization problems, employing parameterized quantum circuits to search for the optimal solutions to these problems. As a result, there are significant differences in design principles and application domains between HVA and QAOA.

The Symmetry Breaking HVA~\cite{babbush2018low} aims to effectively reduce the complexity of quantum circuits by leveraging symmetries inherent in the problem. By exploiting the symmetries of the problem, it disrupts the system's symmetry to represent better and handle the problem's characteristics, thereby designing more suitable quantum circuit structures. The Fourier-transform HVA~\cite{vogt2020preparing} introduces the concept of Fourier transformation. By transforming the energy function of the problem using Fourier transformation, it shifts the problem domain to the frequency domain, making the characteristics of the problem more apparent in the frequency domain. Consequently, it facilitates the design of quantum circuit structures tailored to the problem's characteristics.
\end{itemize}

%QCE
\subsection{Quantum Circuit Embedding (QCE)}
\label{QCE}
\begin{itemize}[leftmargin=1.5em]

\item \textbf{Intent}:
Increasing large-scale unstructured data, such as in knowledge graphs and QML, leads to slow training and inductive inference. By embedding classical information in quantum circuits, the quantum circuit embedding model can open up a more comprehensive range of applications for quantum computing and facilitate the integration of classical and quantum computing.

\item \textbf{Motivation}:
An essential issue for quantum learning algorithms is efficiently generating quantum states from classical data. QCE considers only real-valued representations stored in the classical data structure $T$. Then, by accessing the classical data structure $T$, quantum states corresponding to potential features can be efficiently generated. QCE realizes the quantum ansatz through four building blocks consisting of variational or controlled gates. The choice and parameters of these quantum gates can be adapted to the properties of the classical problem for efficient encoding and processing of classical information.

\item \textbf{Applicability}:
QCE is a versatile tool for transforming problems from various domains into quantum states and solving them on quantum computers.

\item \textbf{Circuit Diagram}:
Please refer to Figure~\ref{fig:QCE}.

\begin{figure}[h]
  \centering
  \resizebox{0.8\linewidth}{!}{
   \begin{quantikz}[row sep=0.7cm,wire types={q,q,q,q}]%
    %qubit 0
    &\gate{R_x(x_1)} &\gate[2]{ZZ(\theta_1)} & &\gate[4]{ZZ(\theta_1)} &\gate{R_Y(\theta_5)}&\\
    %qubit 1
    &\gate{R_x(x_2)} & &\gate[2]{ZZ(\theta_3)} & &\gate{R_Y(\theta_6)}&\\ 
    %qubit 2
    &\gate{R_x(x_3)} &\gate[2]{ZZ(\theta_2)} & & &\gate{R_Y(\theta_7)}&\\
    %qubit 3
    &\gate{H} & & & &\gate{R_Y(\theta_8)}&
    \end{quantikz}
    }
  \caption[Quantum Circuit Embedding (QCE)]{Quantum Circuit Embedding (QCE) ansatz for a single trainable layer~\cite{lloyd2020quantum}}
  \label{fig:QCE}
\end{figure}
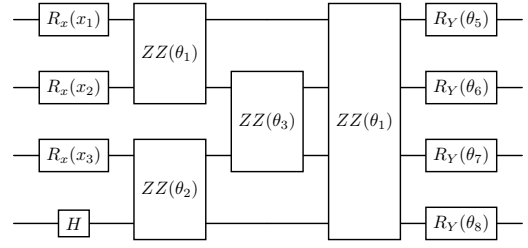

\item \textbf{Example}:
Lloyd et.al~\cite{lloyd2020quantum} proposed a hybrid quantum-classical embedding using QCE ansatz to replace conventional classifiers. Realized binary image classification based on PennyLane. 

\item \textbf{See Also}:
The QCE approach can be used to encode conventional data into quantum states and applied in QML. In QCE, the structure of the quantum circuit is fixed, but the parameters within the circuit are adjustable. However, in Fully Parameterized Quantum Circuit Embedding ({\scriptsize F}QCE)~\cite{ma2019variational}, not only is the structure of the circuit adjustable but the topology of the circuit can also be freely modified, making it more flexible. As a result, {\scriptsize F}QCE offers a higher degree of parameterization and flexibility, allowing it to better adapt to various quantum computing tasks and hardware architectures. \black{Additionally, to enhance the quality of embedded data, Quantum Embedding Kernels (QEKs)~\cite{hubregtsen2022training} address the challenge of selecting an appropriate kernel function by employing the kernel-target alignment method.}

%Additionally, Quantum Embedding Kernels (QEKs)~\cite{hubregtsen2022training} proposed the use of kernel-target alignment, increase the quality of embedded data.
\end{itemize}

\subsection{Multiscale Entanglement Renormalization Ansatz (MERA)}
\label{MERA}
\begin{itemize}[leftmargin=1.5em]

\item \textbf{Intent}: 
The key to efficient algorithms for the numerical simulation of quantum many-body systems is using tensor networks to represent them. It is essential to efficiently compute the expected value of the local serviceability from the tensor network. Therefore, the MERA has been proposed to efficiently encode the quantum many-body state of a D-dimensional lattice system and accurately compute its local expectation.

\item \textbf{Motivation}: 
The MERA is a representation for describing quantum states in quantum many-body systems. It is based on the entanglement structure of quantum information and characterizes the system's entanglement by a series of tensor network layers. In MERA, each layer represents the quantum entanglement in the system at different scales, from micro to macro. This hierarchical structure allows MERA to capture the quantum entanglement information at all scales in the system, providing an effective way to study the nature and behavior of quantum many-body systems.

\item \textbf{Applicability}: 
For quantum many-body states on a D-dimensional lattice, where one wants to accurately and efficiently compute the expectation value of a local observable. It is also the basis for entanglement renormalization, a new coarse-graining method for quantum many-body systems on lattices.

\item \textbf{Circuit Diagram}: 
Please refer to Figure~\ref{fig:MERA}.

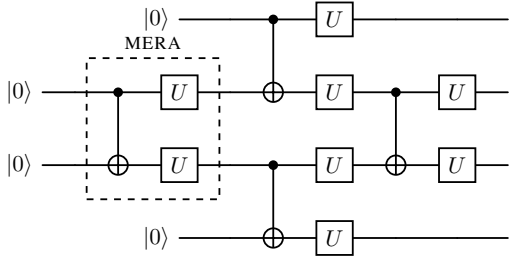
\begin{figure}[h]
  \centering
  \resizebox{0.8\linewidth}{!}{
    \begin{quantikz}[row sep=0.6cm,wire types={n,q,q,n}]%
    %qubit 0
    &&&\lstick{\ket{0}} & \setwiretype{q}&  \ctrl{1} & \gate{U} &&&\\
    %qubit 1
    \lstick{\ket{0}}&&   \ctrl{1}\gategroup[2,steps=2,style={dashed, inner xsep=6pt}]{{\sc mera}}  & \gate{U} && \targ{} &\gate{U} &\ctrl{1} & \gate{U}  & \\ 
    %qubit 2
    \lstick{\ket{0}}&&  \targ{} & \gate{U}  && \ctrl{1} & \gate{U}& \targ{} &\gate{U} & \\
    %qubit 3
    &&&\lstick{\ket{0}} & \setwiretype{q}  &\targ{} &\gate{U}&&&
    \end{quantikz}
}
  \caption[MERA]{Multiscale Entanglement Renormalization Ansatz (MERA)~\cite{vidal2008class}}
  \label{fig:MERA}
  \vspace*{-0.2cm}
\end{figure}

\item \textbf{Example}: 
Rizzi {\it et al.}~\cite{rizzi2008simulation} proposed an algorithm to simulate time evolution using MERA. Montanaro~\cite{montanaro2016quantum} demonstrated a method for computing critical exponents of one-dimensional quantum critical systems based on an iterative scheme applied to MERA.

\item \textbf{See Also}: 
MERA exponentially increases the circuit depth by adding initial qubit $|0\rangle$. Reversed MERA reduces the circuit depth. In quantum machine learning, QCNN leverages this property as a convolution-pooling layer.
\end{itemize}

\subsection{Quanvolutional Neural Network (QNN)}
\label{QNN}
\begin{itemize}[leftmargin=1.5em]

\item \textbf{Intent}: 
CNN has become the standard for many machine-learning applications. However, classical computing still has problems that CNN cannot solve, such as complex problems in pattern recognition, optimization, machine learning, and other areas. Introducing quantum transformations to CNN may leverage quantum advantages to help solve these problems.

\item \textbf{Motivation}: 
QNN is a hybrid classical-quantum algorithm. It extends CNN by replacing classical convolutional layers with quantum convolutional layers. The quantum layer consists of multiple quantum filters that encode input data into quantum states and use random quantum circuits to extract features. By optimizing the evolution process of quantum states, QNN can perform tasks such as classification and regression of input data.

\item \textbf{Applicability}: Levering some form of non-linear quantum circuit transformations for machine learning purposes.

\item \textbf{Circuit Diagram}: 
Please refer to Figure~\ref{fig:QNN}.

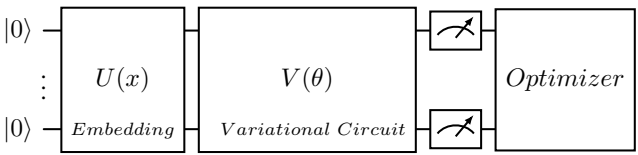
\begin{figure}[h]
    \resizebox{1.0\linewidth}{!}{
    \centering
    \begin{quantikz}[row sep=0.1cm,column sep=0.2cm,wire types={q,q,q,q}]
    %qubit 0
    \lstick{\ket{0}}& \gate[3][1.7cm]{U(x)} &\gate[3][3cm]{V(\theta)} &\meter{} &\gate[3]{Optimizer}\\ 
    %qubit 1
    \vdots \setwiretype{n} \\
    %qubti 2
    \lstick{\ket{0}}& \gateoutput{$Embedding$} &\gateoutput{$Variational\  Circuit$} &\meter{} &
\end{quantikz}
}
  \caption[fig:QNN]{Quanvolutional Neural Network (QNN)~\cite{henderson2020quanvolutional}}
  \label{fig:QNN}
  
\end{figure}

\item \textbf{Example}: Yang {\it et al.}~\cite{yang2021decentralizing} proposed a novel decentralized feature extraction method in joint learning based on QNN to solve the privacy preservation problem in speech recognition. Oh {\it et al.}~\cite{oh2021quantum} use superposition and entanglement to store large amounts of data logarithmically in qubits. QNN model to efficiently process massive data.

\item \textbf{See Also}: 
QNN provides an early real-world application for quantum machine learning.  
\end{itemize}

\subsection{Quantum Convolutions Neural Network (QCNN)}
\label{QCNN}
\begin{itemize}[leftmargin=1.5em]

\item \textbf{Intent}: 
Using machine learning and computational methods to analyze exponentially complex quantum many-body systems.

\item \textbf{Motivation}: 
Large-scale neural networks have solved classical image recognition and error correction problems. Classical machine learning faces many challenges in solving quantum many-body problems due to the vast many-body Hilbert space.
QCNN provides a machine-learning approach based on quantum many-body physics to address these challenges. The convolution layer applies a single quasilocal unitary in a translationally invariant manner for finite depth. Based on the measurement results, the pooling layer measures a portion of the qubit and influences nearby qubits. The convolution and pooling layers are iterated alternately until the system size is sufficiently small. A fully connected layer obtains the result. 

\item \textbf{Applicability}: 
Using realistic near-future quantum devices for efficient machine learning. %解决quantum phase recognition(QPR)以及quantum error correction (QEC) optimization
Addressing Quantum Phase Recognition (QPR) and Optimizing Quantum Error Correction (QEC).

\item \textbf{Circuit Diagram}: 
Please refer to Figure~\ref{fig:QCNN}.

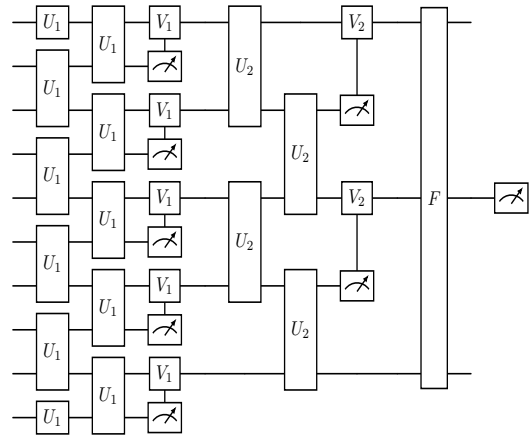
\begin{figure}[h]
  \centering
  \resizebox{0.8\linewidth}{3cm}{
    \begin{quantikz}[row sep=0.2cm,column sep = 0.5cm]
    %qubit 0
    & \gate{U_1} &\gate[2]{U_1} &\gate{V_1} &&\gate[3]{U_2} & &\gate{V_2} &&\gate[9]{F} &\\
    %qubit 1
    & \gate[2]{U_1} & &\meter{}\wire[u]{q} \\
    %qubit 2
    & & \gate[2]{U_1} &\gate{V_1} && &\gate[3]{U_2} &\meter{}\wire[u][2]{q}\\
    %qubit 3
    & \gate[2]{U_1} & &\meter{}\wire[u]{q}\\
    %qubit 4
    & & \gate[2]{U_1}  &\gate{V_1} &&\gate[3]{U_2} & &\gate{V_2} && &&\meter{}\\
    %qubit 5
    & \gate[2]{U_1} & &\meter{}\wire[u]{q}\\
    %qubit 6
    & &\gate[2]{U_1}  &\gate{V_1} & &&\gate[3]{U_2} &\meter{}\wire[u][2]{q} \\
    %qubit 7
    & \gate[2]{U_1} & &\meter{}\wire[u]{q}\\
    %qubit 8
    & &\gate[2]{U_1} &\gate{V_1} & &&&&&& \\
    %qubit 9
    & \gate{U_1} & &\meter{}\wire[u]{q} 
    \end{quantikz}
    }
  \caption[fig:QCNN]{Quantum Convolutional Neural Network (QCNN)~\cite{cong2019quantum}}
  \label{fig:QCNN}
  \vspace*{-0.2cm}
\end{figure}

\item \textbf{Example}: Cong {\it et al.}\cite{cong2019quantum} explicitly described how QCNNs could be used to detect quantum states associated with one-dimensional symmetry-protected topological phases accurately. Herrmann {\it et al.}\cite{herrmann2022realizing} realized a QCNN on a 7-qubit superconducting quantum processor and identified the symmetry-protected topological (SPT) phase of the spin model characterized by a non-zero string order parameter.

\item \textbf{See Also}: 
For any given state $|\phi|$, there is always a QCNN that detects $|\phi|$ with a deterministic measurement result, and such a QCNN can be described as a circuit in the opposite direction to an MERA circuit. This additional degree of freedom distinguishes QCNNs from MERA.
\end{itemize}

\subsection{Assessing Support for Ansatzes in Existing Quantum Computing Platforms}

We preliminary investigated several widely used open-source quantum computing platforms and environments to assess their support for ansatzes. The platforms we examined include IBM Quantum~\cite{ibmQuantumComputing}, Google Quantum AI~\cite{quantumaiGoogleQuantum}, HiQ~\cite{huaweicloudHiQx91CFx5B50x8BA1x7B97}, PennyLane~\cite{pennylanePennyLane}, Paddle Quantum~\cite{githubGitHubPaddlePaddleQuantum}, Classiq~\cite{classiqCreateQuantum}, Origin Quantum~\cite{originqcx672Cx6E90x91CFx5B50x8BA1x7B97x5B98x7F51}, and NVIDIA CUDA-Q~\cite{nvidiaQuantumComputing}. The results are summarized in Table~\ref{table:ansatz-2}.

Our findings indicate that each platform offers a variety of ansatz options, ranging from standard choices like UCC and QAOA to more specialized and customizable approaches such as ADAPT VQE. Furthermore, some platforms provide dedicated ansatz libraries with built-in support for specific problem domains, such as quantum chemistry or combinatorial optimization. Overall, the availability and flexibility of ansatz choices vary among platforms, enabling users to select the most suitable option for their particular quantum computing tasks and requirements.

%--------------------
%english table

\begin{table}[t]
% 表设置于文本中央
\centering
\caption{Quantum Computing Platforms and the Ansatzes They Support}
\label{table:ansatz-2}
\renewcommand\arraystretch{0.95}
\footnotesize
\begin{tabular}{|l|p{5cm}|} %l(left)居左显示 r(right)居右显示 c居中显示
\hline 
\multicolumn{1}{|c|}{\textbf{Platform Name}} & \multicolumn{1}{c|}{\textbf{Support Ansatz}} \\ \hline \hline

         IBM Quantum\cite{ibmQuantumComputing} & UCCSD, QAOA, ADAPTV VQE \\ \hline
         Google Quantum AI\cite{quantumaiGoogleQuantum} & UCCSD, QAOA\\ \hline
         HiQ\cite{huaweicloudHiQx91CFx5B50x8BA1x7B97} & HEA, QCC, QAOA, QNN  \\ \hline 
         PennyLane\cite{pennylanePennyLane} & UCCSD, QAOA, MERA, QNN  \\ \hline 
         Paddle Quantum\cite{githubGitHubPaddlePaddleQuantum} & QAOA, QNN \\ \hline
         Classiq\cite{classiqCreateQuantum} & HEA, UCC, QNN \\ \hline 
         Origin Quantum\cite{originqcx672Cx6E90x91CFx5B50x8BA1x7B97x5B98x7F51} & HEA, UCCSD, QAOA \\ \hline
         NVIDIA CUDA-Q\cite{nvidiaQuantumComputing} & HEA, UCCSD, QNN \\ \hline
    \end{tabular}
    \vspace*{-0.3cm}
\end{table}
%--------------------

\section{Discussions}
\label{sec:discussion}

Quantum circuit ansatzes have shown great potential in various quantum computing applications, but the field is still evolving. Several open challenges and exciting opportunities exist for future research. This section will discuss the current limitations, areas for improvement, and potential avenues for further exploration to advance the design and implementation of quantum circuit ansatzes.

\subsection{Current Limitations and Areas for Improvement}

Although quantum circuit ansatzes have shown promise, limitations still need to be addressed.

\begin{itemize}[leftmargin=1.5em]

\item \textbf{Quantum circuit depth:} The circuit depth of ansatzes directly affects their resource requirements and the size of problems they can efficiently handle. Hardware-efficient ansatzes, tailored for near-term quantum devices with limited qubit and gate resources, often lack the expressiveness needed to explore larger problem spaces. Future research should focus on designing ansatzes that balance hardware efficiency and expressive power, enabling the efficient exploration of complex quantum problems.
%and increasing the quantum volume - a crucial measure of quantum computational capacity.

\item \textbf{Error mitigation and noise resilience:} Quantum computing platforms are prone to errors and noise, which can negatively affect the accuracy and reliability of quantum algorithms. Developing ansatzes with built-in error mitigation techniques, such as error-correcting codes and error-robust quantum circuits, is vital to enhance the resilience of quantum algorithms in practical settings. Addressing error mitigation challenges will be instrumental in harnessing the true potential of quantum circuit ansatzes in noisy quantum hardware.

\item \textbf{Ansatz adaptability and interoperability:} As quantum hardware continues to evolve rapidly, the adaptability and interoperability of quantum circuit ansatzes become essential. Researchers must design ansatzes that can dynamically adjust to varying hardware constraints and capabilities, allowing seamless execution of quantum algorithms across different quantum computing platforms. Improving interoperability between different quantum devices and ansatzes will increase accessibility and implementation of quantum algorithms.

\item \textbf{Optimization efficiency:} Optimizing ansatz parameters is computationally intensive, especially for large-scale problems. To improve the efficiency of ansatz optimization, future research should explore advanced optimization techniques, such as machine learning-inspired approaches and gradient-free optimization methods. Leveraging machine learning to accelerate parameter tuning can significantly reduce the computational burden and facilitate the practical implementation of quantum algorithms.

\item \textbf{Quantum-classical hybridization:} Quantum-classical hybrid approaches offer a promising avenue for addressing complex problems efficiently. Combining classical optimization techniques with quantum circuits allows for improved optimization of quantum algorithms and enhances their performance. Exploring novel quantum-classical hybrid ansatzes and investigating their applications in optimization, machine learning, and simulation tasks will pave the way for the next generation of quantum algorithms with enhanced capabilities.

\end{itemize}

\subsection{Potential Issues for Future Research}

Looking into the future, several promising paths exist for research and exploration in quantum circuit ansatz design.

\begin{itemize}[leftmargin=1.5em]

\item \textbf{New circuit primitives and quantum gates:}  Investigating and designing new quantum circuit primitives and specialized gates can introduce new degrees of freedom in ansatz design. By exploring unconventional quantum gates and their impact on circuit expressiveness and efficiency, researchers can uncover hidden efficiencies and enable the development of more powerful quantum algorithms.

\item \textbf{Adaptive ansatz designs:} Developing adaptive ansatz designs that dynamically adjust to hardware constraints and optimize performance for different quantum devices will be pivotal for scaling quantum algorithms on diverse platforms. Adaptive ansatzes will enable quantum algorithms to leverage specific quantum hardware's unique features and capabilities, maximizing performance and efficiency.

\item \textbf{Quantum error-correcting ansatzes:} Integrating error-correcting codes into ansatz designs can significantly enhance fault tolerance and error resilience, making quantum algorithms more robust against noise and errors. Exploring the incorporation of quantum error correction at the ansatz level will be instrumental in advancing practical quantum computing applications.

\item \textbf{Quantum circuit compression techniques:} Quantum circuit compression techniques aim to reduce resource requirements while preserving the expressiveness of quantum circuits. Developing compression techniques that efficiently represent quantum states and circuits will be valuable for reducing qubit and gate overhead, resulting in more resource-efficient implementations of quantum algorithms.

\item \textbf{Automated ansatz selection:} With the increasing complexity of quantum algorithms and the growing number of ansatz options, automated ansatz selection based on problem characteristics and quantum hardware constraints can simplify the process of choosing the most suitable ansatz for a given task. Exploring machine learning-based methods for automated ansatz selection will streamline algorithm design and accelerate the development of new quantum computing applications.

\end{itemize}

\section{Related Work}
\label{sec:related-work}

Design patterns in classical software engineering~\cite{gamma1995design} have been extensively studied and recently extended to quantum computing~\cite{leymann2019towards,buhler2023patterns}. Our work is the first to compile various types of quantum circuit ansatzes from a software engineering perspective, aiming to foster ansatz abstraction and reuse in quantum algorithm design and implementation.

Leymann~\cite{leymann2019towards} proposed the first pattern language for quantum algorithm development, containing ten basic patterns. This work systematized pattern issues, making them a subject of software engineering for quantum algorithms. Subsequent works~\cite{truger2024warm,buhler2023patterns,beisel2022patterns,weigold2020data,weigold2021encoding} have identified and studied new patterns to extend this language. Unlike these studies, our work focuses especially on quantum circuit ansatzes, a special type of design pattern used in designing and implementing common types of VQAs such as VQE~\cite{peruzzo2014variational}, QAOA~\cite{farhi2014quantum}, and QML~\cite{dunjko2016quantum,biamonte2017quantum}. The ansatzes we studied are well-established and used in practical quantum algorithm design. Our aim is not to propose new patterns (ansatzes) but to create a catalog of quantum circuit ansatzes to support their abstraction and reuse in practical quantum algorithm design. Recently, Perez-Castillo {\it et al.}~\cite{perez-castillo-24-patterns} conducted an empirical study analyzing the application of five quantum design patterns proposed in~\cite{leymann2019towards}: initialization, uniform superposition, oracle, entanglement, and uncompute. This study helps quantum software developers understand how and when to apply these design patterns.

Some research has summarized quantum circuit ansatzes for various purposes, such as the work by Tilly {\it et al.}~\cite{tilly2022variational} discussing ansatz selection for the VQE algorithm pipeline and the overview of VQAs by Cerezo {\it et al.}~\cite{cerezo2021variational}. \black{Also, some works~\cite{gu2024pseudomagic,wille2010synthesis,sanchez2023operating,wille2022decision}, although they do not specifically formulate their results as patterns (ansatzes), provide knowledge and insights that can be used to construct better ansatzes (circuits). This information can be beneficial for the design and selection of suitable ansatzes for quantum algorithm development.} However, unlike our work, these works have not systematically proposed an ansatz catalog for quantum algorithm development from a software engineering perspective.

%Note that the patterns identified in~\cite{leymann2019towards} are in no way meant to be exhaustive, and more patterns should be identified in the future to make the patterns and the pattern language practical and valuable.  

%\red{Many existing works have contributed to advancing quantum technology. Leymann~\cite{leymann2019towards} summarized an architectural pattern language to support quantum algorithm development. Wille~\cite{wille2020efficient} provided an overview of efficient quantum compilation methods. Additionally, research~\cite{glasser2019expressive, meyer2023exploiting} has explored tensor network techniques and symmetries to enhance the expressive power of data in quantum machine learning models.
%For a comprehensive review of VQE and its components, refer to~\cite{tilly2022variational,cerezo2021variational}
%}

%, fostering ansatz abstraction and reuse in quantum algorithm design and implementation, similar to design patterns in classical software engineering~\cite{gamma1995design}.

\section{Conclusion}
\label{sec:conclusion}

Quantum circuit ansatzes have emerged as powerful tools in quantum computing, enabling the design and implementation of quantum algorithms for diverse applications. This paper presents a comprehensive catalog of quantum circuit ansatzes, exploring their unique features and applications. We aim to provide researchers and practitioners in the quantum computing community with a valuable resource for quantum algorithm design and implementation. 

The catalog of quantum circuit ansatzes will help designers design quantum algorithms. By offering a diverse range of ansatzes, researchers now have a new resource to assist them in developing and tailoring quantum algorithms for specific tasks with varying degrees of hardware constraints.

The development and refinement of quantum circuit ansatzes are ongoing endeavors that demand collective efforts from quantum researchers and developers. We call on the quantum computing community to actively engage in the following actions:

\begin{itemize}[leftmargin=1.5em]
\item \textbf{Collaborative Research:} Encourage collaboration and information sharing among researchers working on different ansatzes. The exchange of ideas, code implementations, and benchmarking results can accelerate progress and lead to new promising results in quantum ansatz and circuit design.

\item \textbf{Community Challenges:} Organize community-driven challenges and competitions focused on ansatz optimization and design. These challenges foster innovation and inspire researchers to explore novel ideas and approaches.

\item \textbf{Algorithm Exploration:} Continue exploring the potential of quantum-classical hybrid ansatzes, novel quantum gates, and quantum error-correcting ansatzes to address existing limitations.

\item \textbf{Open-Source Repositories:}
Promote open-source development and maintain repositories that share ansatz implementations, research findings, and empirical studies. Open access to code and data will foster transparency, collaboration, and reproducibility in quantum ansatz and algorithm design.

\item \textbf{Interdisciplinary Collaboration:}
Collaborate with experts in diverse domains such as quantum chemistry, materials science, optimization, and machine learning to explore the practical implications of ansatzes in solving real-world problems.

\end{itemize}
%\newpage
\balance
\bibliographystyle{IEEEtranS}
\bibliography{IEEEabrv,qsw-bibliography}

\end{document}